\documentclass[%
 reprint,
 superscriptaddress,
 amsmath,amssymb,
 aps,
prl,
]{revtex4-2}

\usepackage{dcolumn}
\usepackage{bm}
\usepackage{amsmath}
\usepackage[T1]{fontenc}
\usepackage{xspace}
\usepackage{comment}
\usepackage{bbold}
\usepackage{braket}
\usepackage{ascmac}

\ifx\pdfoutput\undefined
\usepackage[dvipdfmx]{graphicx}
\usepackage[dvipdfmx]{hyperref}
\usepackage[dvipdfmx]{color}
\usepackage[dvipdfmx]{xcolor}
\else
\usepackage{graphicx}
\usepackage[colorlinks=true,linkcolor=blue,anchorcolor=red,citecolor=blue, urlcolor=blue]{hyperref}
\usepackage{color}
\usepackage{xcolor}
\fi

\graphicspath{
{./}
}

\begin{document}

\title{Machine Learning Assisted Characterization of Labyrinthine Pattern Transitions}

\author{Kotaro Shimizu}
\affiliation{Department of Physics, University of Virginia, Charlottesville, Virginia 22904, USA}
\affiliation{Department of Applied Physics, The University of Tokyo, Tokyo 113-8656, Japan} 

\author{Vinicius~Yu~Okubo}
\affiliation{Dept. Electronic Systems Engineering, Polytechnic School, University of São Paulo, Brazil}

\author{Rose Knight}
\affiliation{Department of Physics, University of Virginia, Charlottesville, Virginia 22904, USA}

\author{Ziyuan Wang}
\affiliation{Department of Physics, University of Virginia, Charlottesville, Virginia 22904, USA}

\author{Joseph Burton}
\affiliation{Department of Physics, University of Virginia, Charlottesville, Virginia 22904, USA}

\author{Hae Yong Kim}
\affiliation{Dept. Electronic Systems Engineering, Polytechnic School, University of São Paulo, Brazil}

\author{Gia-Wei Chern} 
\affiliation{Department of Physics, University of Virginia, Charlottesville, Virginia 22904, USA}

\author{B. S. Shivaram}
\affiliation{Department of Physics, University of Virginia, Charlottesville, Virginia 22904, USA}

\date{\today}

\begin{abstract}
We present a comprehensive approach to characterizing labyrinthine structures that often emerge as a final steady state in pattern forming systems. We employ advanced machine learning based pattern recognition techniques to identify the types and locations of topological defects of the local stripe ordering. 
Applying this method to single-crystal Bi-substituted Yttrium Iron Garnet films, we uncover a distinct morphological transition between two zero-field labyrinthine structures.
Crucially, the pair distribution functions of the topological defects reveal subtle differences between labyrinthine structures which are beyond conventional characterization methods. By systematically analyzing the spatial correlations and geometric properties of these defects, we provide new insights into the athermal dynamics governing the observed morphological transitions.
Our work demonstrates that machine learning based recognition techniques enable novel studies of rich and complex labyrinthine type structures universal to many pattern formation systems.  
\end{abstract}

\maketitle

Labyrinthine structures are ubiquitous in out-of-equilibrium nonlinear systems ranging from biological and  chemical reactions to fluid convection, crystal growth, and magnetic ordering~\cite{Nicolis77,Cross93,Koch94,Pismen06,Cross09}. In such pattern forming systems, the complex structures emerge as a result of competing interactions in a highly nonlinear way. The labyrinthine patterns are generally characterized by stripe domains of different orientations, sizes, and grain-boundary structures. The predominance of periodic stripes indicates breaking of translational symmetry locally. Yet, contrary to long-range ordered states in an equilibrium phase transition, labyrinthine patterns are essentially disordered and cannot be described by a well-defined order parameter. Indeed, labyrinthine structures can be viewed as intermediate between a featureless short-range correlated glassy state and long-range ordered stripe or crystalline phases~\cite{Berre02}.  

Despite their prevalence in pattern forming systems, a complete characterization of labyrinthine structures is still lacking~\cite{Alar20}.  A defining characteristic of labyrinthine patterns is the ring-like feature in its structure factor obtained from conventional Fourier analysis~\cite{Elder92}. The radius and width of the ring correspond to the wavelength of local stripes and characteristic size of stripe domains, respectively~\cite{Elder92,Ouyang91,Cross95,Christensen98}. While such global Fourier analysis provides a basic characterization of labyrinths, it fails to capture subtle differences of labyrinthine patterns which have important structural or dynamical implications. Other useful measures, such as the disorder functions~\cite{Gunaratne95,Hu04,Hu05}, have been introduced to quantify deviations from a perfect stripe order. Another important characterization often employed is the density of topological defects of labyrinthine structures~\cite{Hou97,Seul92,Seul92b,Seul92c}. Indeed, the distribution and correlation between topological defects, such as disclinations and dislocations, of the stripe order encode important information about the labyrinths~\cite{Seul92b,Seul92c}. However, efficient and accurate identifications of such point-like defects in large-scale experimental or simulation data remain a challenging task.

In this paper, we present a comprehensive framework for the characterization of labyrinthine structures by leveraging machine learning based template recognition methods. 
A two-step algorithm that consists of rotation-invariant template matching followed by convolutional neural network analysis was developed to precisely identify topological defects (junctions and terminals) and their coordinates in the images. The high-precision real-space configuration data allow us to compute pair distribution functions of the topological defects, which provide valuable information about labyrinthine patterns that are complementary to those extracted from structure factors. 

We apply our approach to studying an intriguing nonequilibrium morphological phase transition of domains in Ytrium Iron Garnet (YIG), a well known technologically important ferromagnetic material. Magnetic interactions in YIG are dominated by a short-range ferromagnetic exchange and long-range dipolar interactions~\cite{Hansen84}. YIG films doped with bismuth are known to carry a strong perpendicular magnetic anisotropy and form complex labyrinthine patterns ~\cite{Seul92,Seul92b,Seul92c}. Bismuth doping introduces a number of advantages desirable for magneto-optic isolators and sensors~\cite{Fratello96,Fratello86}, enhances the Faraday rotation, improves the perpendicular anisotropy and also lowers the saturation magnetization~\cite{Hansen84,Fratello96,Fratello86,Fakhrul19}.  For films grown under appropriate conditions, the magnetization can be saturated in fields less than 100 Oe.  Thus the labyrinthine stripe domain patterns are easily observed with a small electromagnet in polarized light under a microscope~\cite{Puchalska78}.

\begin{figure}[tb]
\centering
\includegraphics[width=0.85\columnwidth]{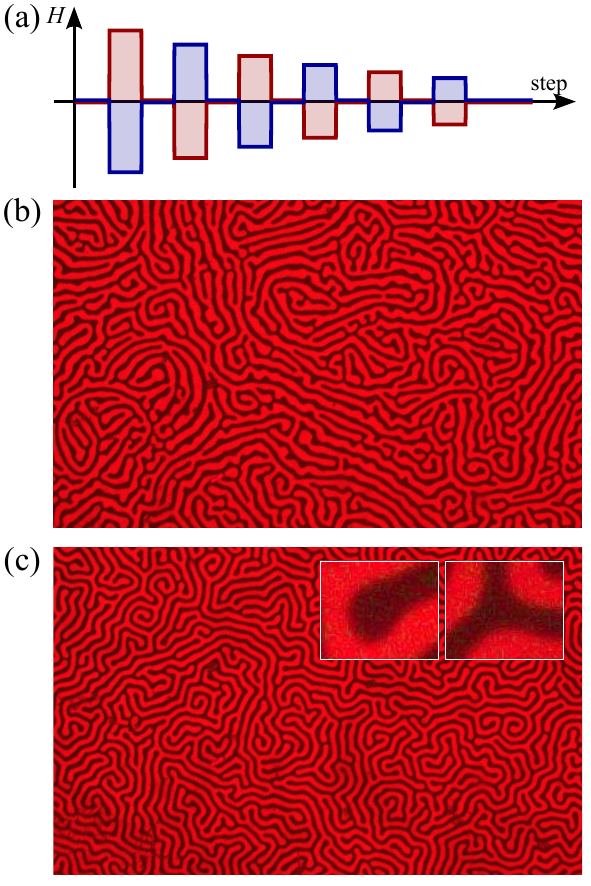}
\caption{
\label{fig:labyrinth}
(a) Schematic figure of the external magnetic field stepping protocol. The sequence starting from the positive (negative) magnetic field is denoted by the red (blue) line.
Domain images of (b) the quenched state at step 0 and (c) the annealed state at step 36. 
The left and right insets in (c) display the structures of a terminal and a junction, respectively. 
}
\end{figure}

Specifically, here we report an intriguing pattern transition in experiments performed on YIG films where a perpendicular magnetic field is stepped down to zero starting from a fully saturated state following an ``annealing'' protocol. In this protocol starting from the sample in the fully magnetized state, we instantly drop the field to zero where it is held for 10 seconds during which an image is acquired. Further sequential de-magnetization is carried out but with the magnitude of the field reduced at each step exponentially and with alternate reversal of field directions as indicated in Fig.~\ref{fig:labyrinth}(a). All measurements reported here were performed at room temperature.  The images obtained covered an area of 2 mm $\times$ 1.8 mm.  Two samples grown under similar conditions were studied.\\

Representative images obtained during the beginning of the protocol and at the end of the protocol are shown in Fig.~\ref{fig:labyrinth}(b) and Fig.~\ref{fig:labyrinth}(c), respectively.
Both of them exhibit the labyrinthine stripe patterns and consist of a plethora of defects depicted in Fig.~\ref{fig:labyrinth}(c).
These defects where the dark domain ends and three domains meet are termed terminals ($+1/2$ disclinations) and junctions ($-1/2$ disclinations), respectively.  These are the topological defects associated with rotational symmetry breaking~\cite{Seul92}. When introducing one of these defects into perfectly ordered stripes, it inevitably gives rise to the appearance of the other. In other words, the disorder manifests as a disclination pair.

\begin{figure}[b]
\centering
\includegraphics[width=0.9\columnwidth]{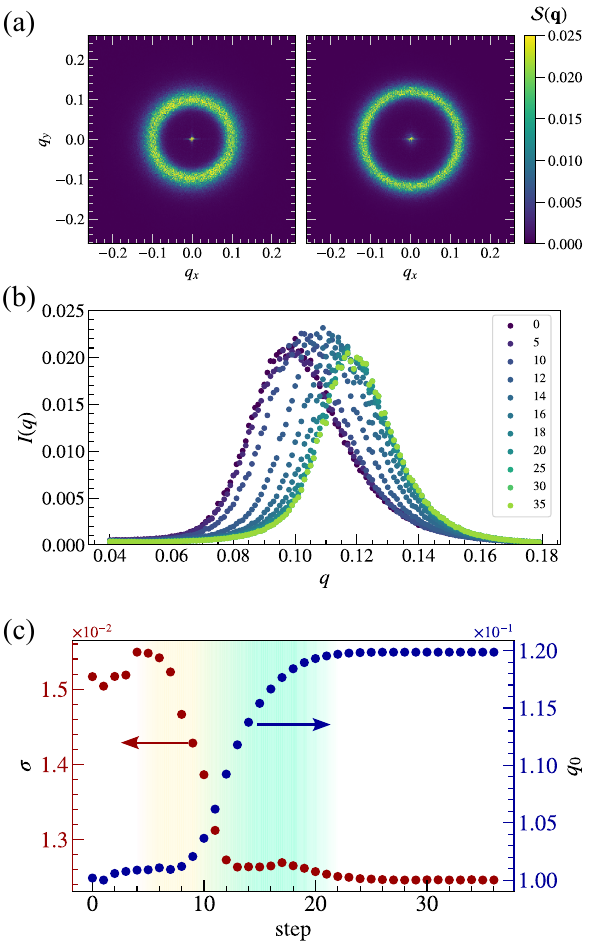}
\caption{
\label{fig:fourier}
(a) Structure factor of the experimentally-obtained images of labyrinthine patterns $\mathcal{S}({\bf q})$ in Eq.~\eqref{eq:structure_factor} for (left) the quenched state at step $0$ and (right) the annealed state at step $36$.
(b) The angle-averaged structure factor, $I(q)$, obtained by averaging $\mathcal{S}(\mathbf{q})$ over the polar angle in the Fourier space, for different steps represented by different colors. 
(c) Evolution of the width of the peak in $I(q)$, $\sigma$, and its peak wave number, $q_0$, respectively represented by the red and the blue circles. 
}
\end{figure}

\begin{figure*}[tb]
\centering
\includegraphics[width=1.8\columnwidth]{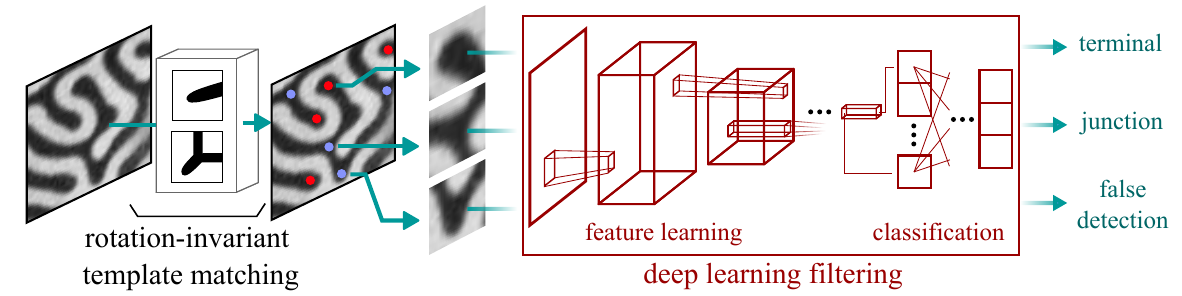}
\caption{
\label{fig:algorithm}
The schematic illustration of the algorithm to detect the defects. 
First, we detect candidates for the defects (red and blue dots) by using the rotation-invariant template matching algorithm. The identified candidates in this process, while many are terminals and junctions, also encompass structures that are not defects. Then, these candidates are classified into terminal, junction, and false detection through the deep learning filter. 
}
\end{figure*}

The two (both in zero field) images show labyrinthine structures of similar domains, but appear visually different at the same time, making it challenging to quantify their morphology. For further analysis and discussion, we distinguish the domain patterns in Figs.~\ref{fig:labyrinth}(b) and \ref{fig:labyrinth}(c) as the quenched and annealed states, respectively. In the quenched state, the border of bright and dark domains exhibit a sinuous nature and do not appear as parallel.  In contrast, the annealed state consists of significant regions of very parallel domains. This state exhibits roughly equal widths of dark and bright domains, and the areas occupied by them are also roughly equal for any given sample region. 
This net approximate equality can be achieved by a variety of different patterns which more or less appear the same.

To characterize these morphologically distinct states, we first perform the conventional Fourier analysis by calculating the structure factor 
\begin{eqnarray}
    \mathcal{S}({\bf q})=
    \Braket{
    \left|w_{\epsilon,n,m}({\bf q})\right|^2
    }_{\epsilon,n}, 
	\label{eq:structure_factor}
\end{eqnarray}
where $w_{\epsilon, n, m}({\bf q})$ is the Fourier transform of the gray-scaled magnetic domain image obtained in experiments with the wave vector $\mathbf{q}=(q_x,q_y)$ at step $m$ ($m=0,1,\ldots,36$) of the $n$-th trial ($n=1,2,\ldots,6$) under the field stepping protocol starting from the positive ($\epsilon=1$) and negative ($\epsilon=-1$) field, and $\braket{\cdots}_{\epsilon,n}$ represents the average over $\epsilon$ and $n$. 
In Fig.~\ref{fig:fourier}(a), we show the structure factors, manifesting circular patterns due to the isotropic nature of the system.  The circular pattern is smaller and thicker in the quenched state than in the annealed state. 
Since the intensities of the structure factor shown in Fig.~\ref{fig:fourier}(a) are rotationally symmetric, we further computed the angle-averaged structure factor $I(q)$; we introduce the amplitude of the wave vector $q$ and the polar angle $\phi$ as $\mathbf{q}=q(\cos\phi, \sin\phi)$, and average $\mathcal{S}(\mathbf{q})$ over $\phi$ to obtain $I(q)$.
The evolution of $I(q)$ with demagnetization steps are shown in Fig.~\ref{fig:fourier}(b). 
As discussed above, the demagnetization brings about the larger radius of the circular pattern, which becomes apparent as the wave number $q$ where $I(q)$ takes the maximum value moves to the right. 

To capture the intricate evolution of the peak structure, we fit $I(q)$ by a Gaussian given by $A\exp\left(-(q-q_0)^2/\left(2\sigma^2\right)\right)$ and extract the peak width $\sigma$ as well as the peak wave number $q_0$.
The step dependence of $\sigma$ and $q_0$ are shown in Fig.~\ref{fig:fourier}(c), exhibiting the decrease of $\sigma$ and the increase of $q_0$ with steps.
Notably, $\sigma$ and $q_0$ exhibit sharp transitions at distinct steps, specifically, in the range of 6 to 12 steps represented by the yellow 
area and 10 to 20 steps represented by the green area, 
respectively. 
It is also clearly seen that between steps 10 and 12 (when the quenching magnetic field falls below 12 gauss),
the domains settle down to the annealed state.  
Considering that $1/\sigma$ and $2\pi/q_0$ correspond to the correlation length and the magnetic modulation period, respectively, we can infer that through the demagnetization process, the labyrinthine structure, which is initially less compact in the quenched state, aligns itself to increase the correlation length and subsequently shortens the magnetic period, transitioning into a more compact annealed state. 
Application of further smaller steps reorganizes the domain pattern locally at various points but leaves the overall features intact.

The conventional analysis conducted thus far provides a global picture of domain reorganization during the annealing process. However, it fails to yield a comprehensive understanding of the detailed response of topological defects. As the evolution of the labyrinth patterns is mediated by the defects introduced in Fig.~\ref{fig:labyrinth}, the precise identification of their number and coordinates is inevitable for further clarification. 

Today, we have high-resolution digital cameras and much greater computational power compared to what was used in the pioneering work of decades ago~\cite{Seul92,Seul92b,Seul92c}. 
In addition, modern popular algorithms for object detection based on convolutional neural networks~\cite{Lecun1995}, such as Faster R-CNN~\cite{ren2016faster} and YOLO~\cite{redmon2016look}, are available.
However, they are not well-suited for defect detection for two reasons. 
First, they are designed to detect a few large objects, while our goal is to identify thousands of small, closely clustered objects.
Second, they require a large number of manually annotated training images, which is extremely labor-intensive to produce due to the high number of defects in each image.

We developed a two-step detection algorithm capable of identifying thousands of small objects with minimal manual annotation, as shown in Fig.~\ref{fig:algorithm}.
First, we employ rotation-invariant template matching to identify candidate points for junctions and terminals~\cite{Kim2007,Kim2013}. 
This step is executed with a low threshold set to avoid any false negatives, even if it results in several false positives.
This approach significantly reduces the annotation workload, as we now only need to review the candidate points rather than locating and classifying every defect in the image. We selected some images containing candidate points, manually corrected the false positives, and used them to train a convolutional neural network classifier. Therefore, in the second step, the candidate points from the first stage are processed through the classifier, which differentiates between terminals, junctions, and false detections. Manual verification of numerous images indicates that the algorithm's detection accuracy is nearly 100\%.
More details about the proposed algorithm are provided in Ref.~\cite{okubo2024} and the executable program to demonstrate the automated detection for experimental images is available for download~\footnote{\url{https://github.com/okami361/TM-CNN/releases/tag/Windows_v2.3.7}}.

\begin{figure}[tb]
\centering
\includegraphics[width=0.9\columnwidth]{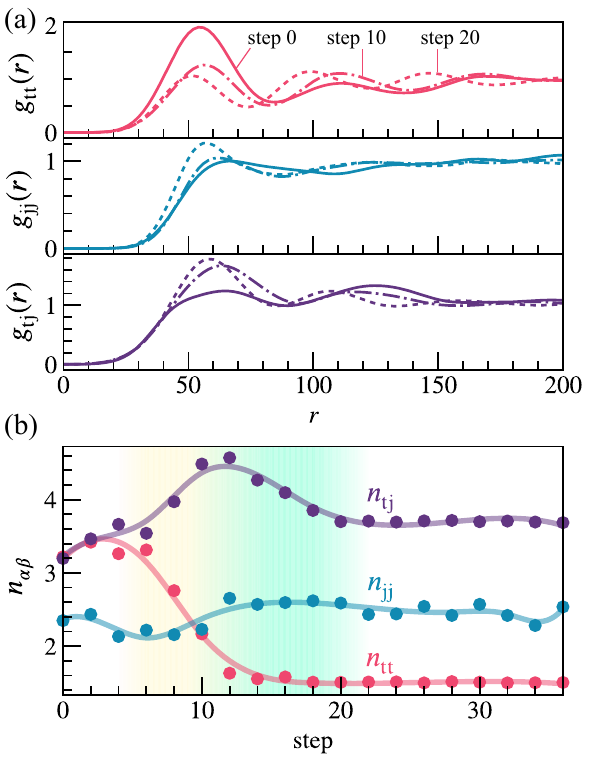}
\caption{
\label{fig:gr}
(a) The radial distribution function $g_{\alpha\beta}(r)$ in Eq.~\eqref{eq:gr} for several steps: $g_{\rm tt}(r)$ (top), $g_{\rm jj}(r)$ (middle), and $g_{\rm tj}(r)$ (bottom). 
The solid, dash-dotted, and dashed lines are for steps 0, 10, and 20, respectively. 
In the calculation of $g_{\alpha\beta}(r)$, the Gaussian convolutional kernel with a variance of 4 is employed for better visibility. 
(b) Evolution of the coordination number $n_{\alpha\beta}$ in Eq.~\eqref{eq:neff} for terminal-terminal pairs (red), junction-junction pairs (blue), and terminal-junction pairs (purple). The pale lines are guides for the eyes. 
}
\end{figure}

To quantify the correlation of topological defects, we calculate the radial distribution function $g_{\alpha\beta}(r)$ for terminals and junctions by
\begin{eqnarray}
    g_{\rm \alpha\beta}(r) = \frac{1}{2\pi r \rho_{\alpha}N_{\rm \beta}}\sum_{ij}\braket{\delta(r- |{\bf r}_{\alpha,i}-{\bf r}_{\beta,j}|)}_n, 
    \label{eq:gr}
\end{eqnarray}
with the detected position of the $i$-th defect $\alpha$ ($\beta$), denoted by $\mathbf{r}_{\alpha,i}$ ($\mathbf{r}_{\beta,i}$); $\alpha$ and $\beta$ can take $\mathrm{t}$ for terminals and $\mathrm{j}$ for junctions. 
In Eq.~\eqref{eq:gr}, the total number and density in the unit of the pixel are respectively represented by $N_{\alpha}$ ($N_{\beta}$) and $\rho_{\alpha}$ ($\rho_{\beta}$).
In Fig.~\ref{fig:gr}(a), $g_{\alpha\beta}(r)$ at steps 0, 10, and 20 are presented. 
Interestingly, $g_{\alpha\beta}(r)$ reveals broad peaks at integer multiples of the magnetic period $2\pi/q_0$, indicating that the disordered features of the defects still exhibit distinct correlation characteristics of each defect. 
The correlation between terminals shown in $g_{\rm tt}(r)$ exhibits a dominant first peak in the quenched state (step 0) with liquid-like correlation, and the peak height diminishes in the annealed state, while higher-order peaks become more pronounced. 
In contrast, for junctions, $g_{\rm jj}(r)$ shows only a barely discernible first peak. 
Unlike terminals, the height of the first peak increases slightly, but overall, the correlation remains much weaker gas-like one. 
In $g_{\rm tj}(r)$, a dominant first peak and a weaker second peak are observed. 
Through the transition from the quenched to the annealed state, the first peak becomes more pronounced, indicating an increase in the number of terminal-junction pairs.

In Fig.~\ref{fig:gr}(b), we illustrate the effective coordination number counting the number of defects $\alpha$ within the nearest-neighboring shell of defects $\beta$ defined by
\begin{eqnarray}
    n_{\alpha\beta} = 
    \int_0^{r_0} 2\pi r \rho_{\rm t} g_{\rm t}(r) dr,
	\label{eq:neff}
\end{eqnarray}
where $r_0$ is the first local minimum of $g_{\alpha\beta}(r)$ around $r \sim 80$. 
Intriguingly, $n_{\alpha\beta}$ exhibits different trends in the demagnetization process depending on the pairs and characterizes the transition from the quenched state to the annealed state by the following two-step process. 
In the initial stage of the transition from step 6 to step 12, highlighted in yellow, terminals significantly reduce the number of neighboring terminals while increasing the number of adjacent junctions. 
Since terminals can pair annihilate with junctions, this stage can be considered a precursor to reducing the total number of defects in the system. 
The increase in paired topological defects enhances the system's coherence, which is consistent with the decrease in $\sigma$ shown in Fig.~\ref{fig:fourier}(c).
After the transition indicated by the peak of $n_{\rm tj}$, terminal-junction pairs annihilate by reducing $n_{\rm tj}$ from step 12 to step 20, highlighted in green. 
Consequently, the space left by the eliminated defects is filled by a denser packing of stripes, leading to an increase in $q_0$, as demonstrated in Fig.~\ref{fig:fourier}(c).

To summarize, we have identified a morphological transition of magnetic domains in a technologically important system and characterized their evolution in new ways by using the topological argument and pattern recognition analysis combined with modern machine learning. The magnetic labyrinthine patterns are notoriously difficult to quantify due to the lack of clear long-range order. Here, by examining the correlations of topological defects, we elucidate how these defects mediate the very subtle transformation from the quenched state to the annealed state and reveal the role of topological defects in the transition. Information about the correlations of such local objects is inaccessible through conventional Fourier analysis and successfully characterizes the transformation of labyrinthine patterns. The systematic detection and correlation analysis of thousands of topological defects has been achieved for the first time through the algorithm we developed.
These versatile tools will undoubtedly find applications in studies of diverse labyrinthine structures.

\begin{acknowledgments}
We thank Dr. Vincent Fratello and Dr. Shanthi Subramanian of Lucent-Bell Labs for the supply of the LPE grown epitaxial thin films used in this work.  We are also grateful to David Huse of Princeton for many helpful comments.  The experimental work was partially funded by the NSF under grant DMR and by the 3Cavalier grant from the University of Virginia. This work was also supported by JSPS KAKENHI Grant Number No. JP21J20812. Gia-Wei Chern was partially supported by the US Department of Energy Basic Energy Sciences under Contract No. DE-SC0020330. K.S. was supported by the Program for Leading Graduate Schools (MERIT-WINGS).
\end{acknowledgments}

\bibliography{ref}

\end{document}